\documentstyle[amssymb,preprint,aps,epsf]{revtex}

\newcommand{\be}{\begin{equation}}
\newcommand{\ee}{\end{equation}}
\newcommand{\bea}{\begin{eqnarray}}
\newcommand{\eea}{\end{eqnarray}}

\begin{document}
\title{Plaquette Ordering in SU(4) Antiferromagnets}
\author{Anup Mishra, Michael Ma, Fu-Chun Zhang,\\}
\address{ Department of Physics,\\
University of Cincinnati, Cincinnati, OH 45221-0011\\
}
\date{\today}
\maketitle


\begin{abstract}
\noindent We use fermion mean field theory to study possible plaquette
ordering in the antiferromagnetic SU(4) Heisenberg model. We find the ground
state for both the square and triangular lattices to be the disconnected
plaquette state. Our mean field theory gives a first order transition for
plaquette ordering for the triangular lattice. Our results suggest a large
number of low lying states.
\end{abstract}

\preprint
\widetext



\widetext

\newpage \narrowtext

Recently, the importance of orbital degeneracy in the physics of the
insulating phase of transition metal oxides has been emphasized \cite
{tokura,kugel,li}. The understanding of the ground state and elementary
excitations in these materials requires consideration of both the orbital
and spin degrees of freedom. In $V_{2}O_{3}$ for example, the anomalous
ordering of the antiferromagnetic insulator can be understood quite simply
as a result of spin-orbital coupling \cite{mcwhan,castellani,bao,park,mila}.
The interplay between spins and orbitals may also be responsible for the
lack of magnetic ordering down to very low temperature in $LiNiO_{2}$ \cite
{takano,kataoka,reynaud,feiner}. The orbital
ordering and correlations have recently been observed in synchrotron x-ray 
diffraction
measurements \cite
{murakami} and in resonant x-ray scattering experiments \cite{paolasini}.

It has been pointed out that for a $S=1/2$ system with a double orbital
degeneracy there is an ideal limit where the spin and orbital degrees of
freedom can be unified into a higher SU(4) symmetry~\cite{arovas,li}.
While this symmetry is usually not realized in the transition metal oxides
due to Hund's rule and anisotropy in hopping amplitudes, the SU(4)
symmetric point may still shed light on the physics of more realistic
systems, especially if the deviation from the ideal limit is not too large.
Using fermion mean field theory(MFT), Li {\it et.al.}~\cite{li} showed that
the SU(4) ground state on a square lattice possesses no magnetic long range
order. Possible candidates for the ground state include the commensurate flux phase
(1/4 flux per plaquette), and the plaquette solid state.
The commensurate flux phase has broken time reversal invariance which is troublesome (although
time reversal invariance may be restored when projected into the physical
Hilbert space of one fermion per site). Simple variational calculations on
the square and triangular lattice support the lack of long ranged order, but
also suggest the ground state should be a plaquette solid or a resonating
plaquette liquid~\cite{li}. On the other hand, Schwinger boson mean field 
theory
predicts a long range ordered state~\cite{shen}. More recently, finite size
numerical calculations on the square lattice provides further evidence that
the ground state has no magnetic long range order \cite{bossche}. However,
while the commensurate flux phase is two-fold degenerate, their numerical
results suggest a 4-fold degeneracy.

A possible explanation for the 4-fold degeneracy is the spontaneous
formation of plaquette state with alternating plaquettes of strong and weak
correlations. This would be consistent with variational calculations. In
this paper, we will investigate plaquette ordering within fermion MFT, a
possibility not considered in the mean field calculation of 
Li {\it et.al.}\cite{li}.
Using fermion MFT, we will show that the uniform bond amplitude mean field
ground state is unstable with respect to plaquette formation on square and
triangular lattices. Plaquette ordering breaks the lattice translational
invariance. Since the symmetry is discrete, a finite temperature phase
transition is possible even in 2 dimensions. Within the MFT, we find the
plaquette transition to be first order for the triangular lattice. For the
square lattice, the result turns out to be ambiguous.

We start with the antiferromagnetic SU(4) Hamiltonian \cite{li}.

\begin{eqnarray}
H =\sum_{\langle{i},{j}\rangle}\sum_{m,n}S^{n}_{m}(i)~S^{m}_{n}(j),
\label{eqn1}
\end{eqnarray}
where $\langle{i},{j}\rangle$ are the pairs of the nearest neighbor sites $i$
and $j$ on the lattice, and $m = 1, 2, 3, 4$ is the flavor index. $S^{n}_m$
are SU(4) generators. The Hamiltonian is equivalent to an isotropic
antiferromagnetic quantum spin-1/2 system with two-fold degenerate orbitals
\cite{li}. In terms of electron operators, $S^{n}_{m}(i)
=~c^{\dagger}_{i,m}c_{i,n}$, where $c_{i,n}$ is the annihilation operator of
an electron of flavor $n$ at site $i$, the Hamiltonian in eq.(\ref{eqn1})
may be rewritten as

\begin{eqnarray}
H = \sum_{\langle{i},{j}\rangle}\sum_{m,n}~c^{\dagger}_{i,m}~c_{i,n}~c^{%
\dagger}_{j,n}~c_{j,m} .  \label{eqn2}
\end{eqnarray}
It is implied that there is a constraint on the number of electrons at every
lattice site, given by $\sum_{n}~c^{\dagger}_{i,n}c_{i,n} = 1$.

In this paper, we shall apply a fermion MFT to study the possibility of
plaquette ordering for the square and triangular lattices. Such an ordering
would break the translational invariance of the crystal lattice. The fermion
MFT is well known from literature~\cite{baskaran,affleck}. We define the
bond operator $\Phi _{ij}=\sum_{n}~c_{i,n}^{\dagger }c_{j,n}$, and introduce
Lagrange multipliers $\lambda _{i}$ to describe the constraint on site $i$.
We then use mean field to approximate the bond operator by its
self-consistently calculated average, and assume that ${\langle \Phi
_{ij}^{\dagger }\rangle }={\langle \Phi _{ij}\rangle }=t_{ij}$ is real.
Also, we replace the local constraint by a global one and set $\lambda
_{i}=\lambda $, independent of the site. The mean field Hamiltonian is then
given by,

\begin{eqnarray}
H_{mf}=-\sum_{\langle {i},{j}\rangle ,n}t_{ij}~\left( c_{i,n}^{\dagger
}~c_{j,n}+h.c\right) +\sum_{\langle {i},{j}\rangle }~t_{ij}^{2}-\lambda
\left( \sum_{i,n}~c_{i,n}^{\dagger }~c_{i,n}-N\right) ,  \label{eqn3}
\end{eqnarray}
where $N$ is the total number of lattice sites, and an overall constant has
been dropped for simplicity. We see that $\lambda $ acts as a chemical
potential.

\begin{figure}[p]
\epsfxsize=6.0in
\begin{center}
\epsffile{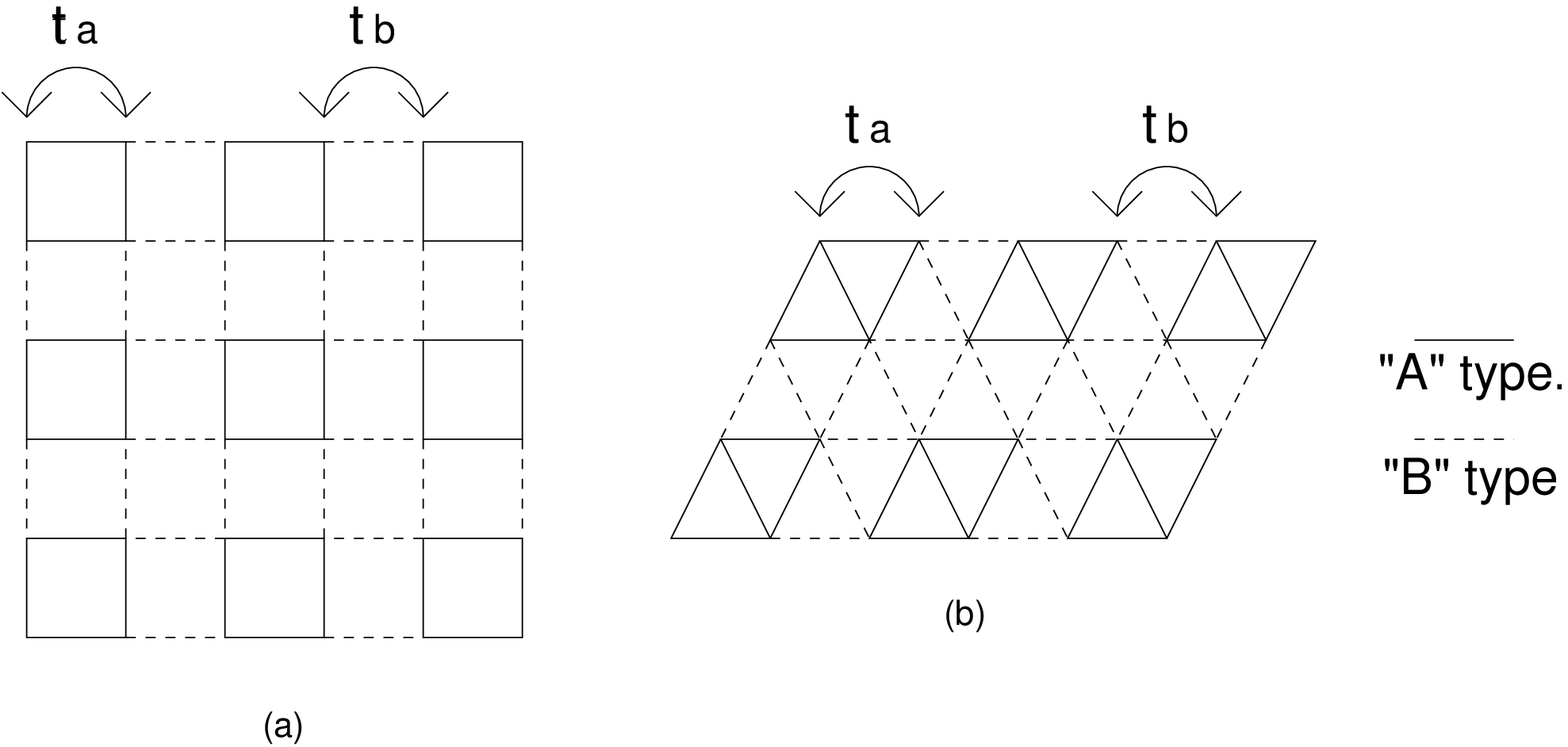}
\end{center}
\caption{{{\ A schematic representation of plaquettes on
a) square lattice, b) triangular lattice. Each solid line represents a
strong bond connecting two lattice sites, with hopping amplitude $t_{a}$.
Each dashed line represents a
weak bond with hopping amplitude $t_{b}$.
}}}
\label{fig1}
\end{figure}

We first discuss the square lattice case. As shown in Fig.(\ref{fig1}), we
consider two types of the bonds in the lattice, denoted by type-A and
type-B, representing stronger and weaker bonds, respectively. In a square
lattice, there are equal numbers of type-A and type-B bonds. The hopping
amplitude on each bond is thus given by

\begin{eqnarray}
t_{ij} = t_{a},~if~\langle{i,j}\rangle \in \{A\}  \nonumber \\
= t_{b},~if~\langle{i,j}\rangle \in \{B\}  \label{eqn4}
\end{eqnarray}
The mean field Hamiltonian becomes
\begin{eqnarray}
H_{mf}= &-&t\sum_{\vec{k},n}\left( \cos {k_{x}a}+\cos {k_{y}a}\right) ~c_{%
\vec{k},n}^{\dagger }c_{\vec{k},n}  \nonumber \\
&+&ir\,t\sum_{\vec{k},n}\left( \sin {k_{x}a}\right) c_{\vec{k}+\left( \pi
,0\right) ,n}^{\dagger }c_{\vec{k},n}  \nonumber \\
&+&ir\,t\sum_{\vec{k},n}\left( \sin {k_{y}a}\right) c_{\vec{k}+\left( 0,\pi
\right) ,n}^{\dagger }c_{\vec{k},n}  \nonumber \\
&+&\frac{Nt^{2}}{2}(1+r^{2}) - \lambda \left( \sum_{\vec{k},n}~c_{\vec{k}%
,n}^{\dagger }c_{\vec{k},n}-N\right).
\label{eqn5}
\end{eqnarray}
In the above eqn., $t=t_{a}+t_{b}$, $r=(t_{a}-t_{b})/(t_{a}+t_{b})$, $%
\vec{k}$ is a crystal momentum, and the sum is over the first Brillouin zone.

In a square lattice, the Hamiltonian is invariant under each of the
following transformations: a) $t\rightarrow -t$, i.e. $(t_{a},t_{b})%
\rightarrow (-t_{a},-t_{b})$; \thinspace b) $r\rightarrow -r$, i.e. $%
(t_{a},t_{b})\rightarrow (t_{b},t_{a})$; and c) $r\rightarrow 1/r$, i.e. $%
(t_{a},t_{b})\rightarrow (t_{a},-t_{b})$. Therefore, it suffices to consider
the parameter space $t\geq 0$, and $0\leq r\leq 1$. The case of $r=0$
corresponds to the uniform bond amplitude mean-field state and the case of $%
r=1$ corresponds to a periodic array of disconnected plaquettes.

The mean field Hamiltonian can be diagonalized and we obtain
\[
H_{mf}=\sum_{\vec{k},\alpha ,n}\epsilon _{\vec{k},\alpha ,n}\beta _{\vec{k}%
,\alpha ,n}^{\dagger }\beta _{\vec{k},\alpha ,n}+\frac{Nt^{2}}{2}\left(
1+r^{2}\right) -\lambda \left( \sum_{\vec{k},\alpha ,n}\beta _{\vec{k}%
,\alpha ,n}^{\dagger }\beta _{\vec{k},\alpha ,n}-N\right) ,
\]
where $\beta _{\vec{k},\alpha ,n}^{\dagger }$ is the creation operator for a
particle in the state $|\vec{k},\alpha ,n\rangle $ with the energy $\epsilon
_{\vec{k},\alpha ,n}$. Here, $\alpha ={1,2,3,4}$ is the band index, and the
sum over $\vec{k}$ is restricted to within the reduced Brillouin zone, which
is a quarter of the first Brillouin zone of the crystal. The reduction of
the Brillouin zone is the result of the translational symmetry broken by the
plaquette state. The four flavor states for a given $\vec{k}$ and $\alpha $
are energetically degenerate. Numerically, we find that for $r\gtrsim 0.3$,
the four energy bands are separated by band gaps. The ground state of the
system is a filled Fermi sea, with Fermi energy $\mu _{f}=\lambda $
determined by the electron number equation,

\[
\sum_{\vec{k},\alpha ,n}\langle \beta _{\vec{k},\alpha ,n}^{\dagger }\beta _{%
\vec{k},\alpha ,n}\rangle =\sum_{\vec{k},\alpha ,n}\theta \left( \lambda
-\epsilon _{\vec{k},\alpha ,n}\right) =N.
\]

\begin{figure}[p]
\epsfxsize=4.5in
\begin{center}
\epsffile{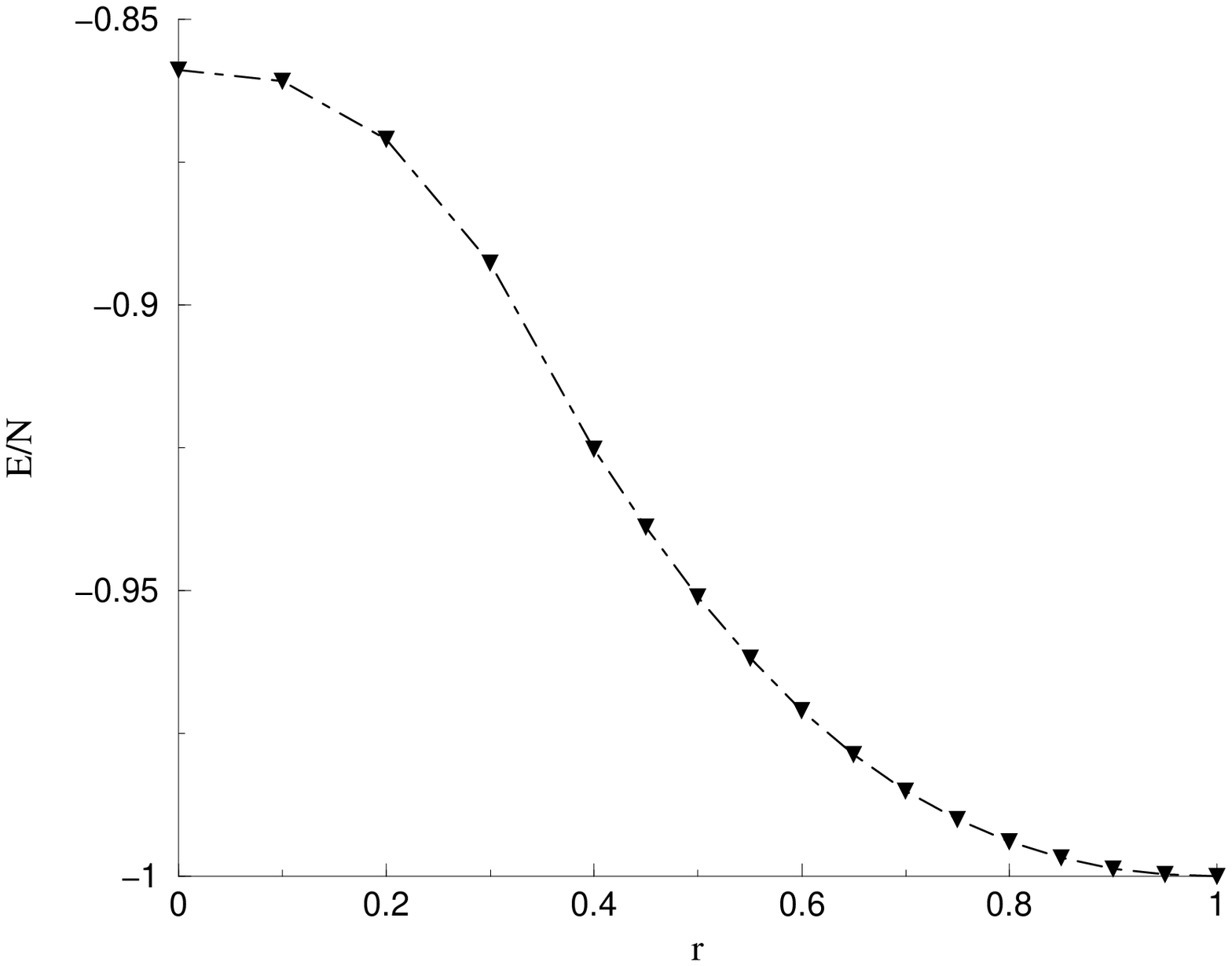}
\end{center}
\caption{{{Energy per site, ($E/N$), in eqn.(\ref{eqn6}) is plotted as a
function of $r$ for a square lattice. $E/N$ monotonically decreases as $r$
increases from $0$ to $1.$ The lowest energy state corresponds to $r=1$, the
disconnected plaquette state.
}}}
\label{fig2}
\end{figure}

To obtain the mean field ground state of the system, we minimize the energy $%
E$ below with respect to $t$ and $r$,
\begin{eqnarray}
\frac{E}{N}=\frac{1}{N}\sum_{\vec{k},\alpha ,n}~\epsilon _{\vec{k},\alpha
,n}\theta \left( \lambda -\epsilon _{\vec{k},\alpha ,n}\right) +\frac{t^{2}}{%
2}\left( 1+r^{2}\right) .
\label{eqn6}
\end{eqnarray}
We first minimize $E$ with respect to $t$ for a given $r$. The energy as a
function of $r$ thus obtained is plotted in Fig.(\ref{fig2}). As shown in
Fig.(\ref{fig2}), the uniform bond amplitude state, given by $r=0$, {\it %
i.e.,} $t_{a}=t_{b}$, is locally unstable with respect to the plaquette
formation. The energy decreases monotonically as $r$ increases from 0 to 1.
The lowest energy state corresponds to $r=1$, the disconnected plaquette
state. The local stability of $r=1$ state can be confirmed independently by
solving the weakly connected plaquette problem using simple second order
perturbation.

The plaquette state is 4-fold degenerate. The discrete symmetry implies the
ordering will remain stable at low temperature in 2D and vanishes either
continuously or through a first order transition at some finite critical
temperature. Within MFT, there will be another critical temperature above
which $t$ vanishes. Since $t$ represents short-ranged order, this transition
is unphysical and is an artifact of this type of MFT.

\begin{figure}[p]
\epsfxsize=4.5in
\begin{center}
\epsffile{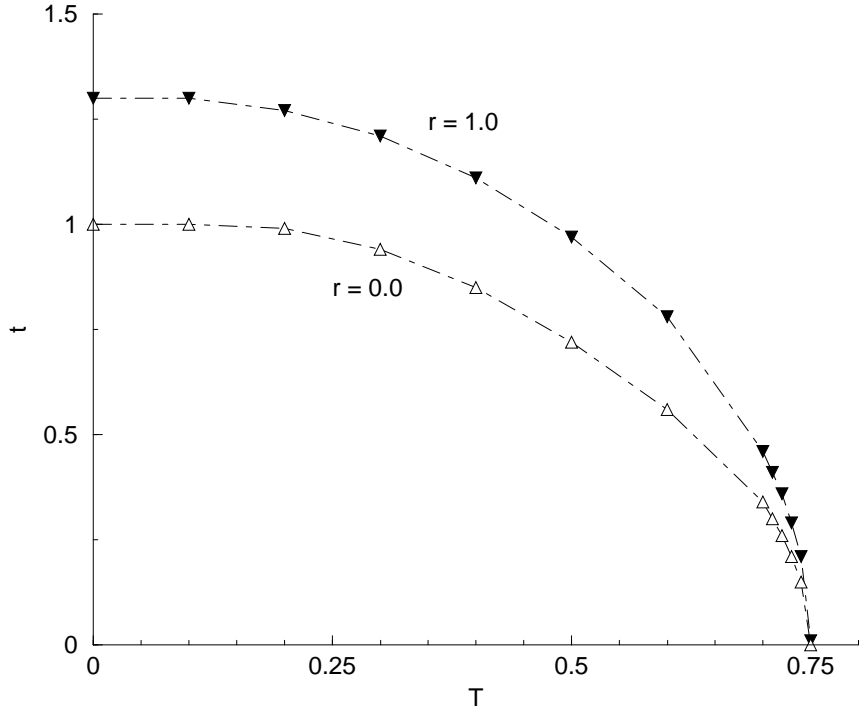}
\end{center}
\caption{{The optimal values of $t$ as functions of temperature $T$ for $r=0$ 
and $r=1$ on a square lattice.
For $ 0 \le r \le 1$, $t$ v.s. $T$ curves lie between these two curves.
The short range correlation
between neighboring lattice sites, characterized by $t$, decreases with
increasing $T$. At $T=0.75$, $t \rightarrow 0$ for all values of $r$.}}
\label{fig3}
\end{figure}

At a finite temperature, the mean field free energy is given by

\[
\frac{F}{N}=-\frac{k_{b}T}{N}\sum_{\vec{k},\alpha ,n}\ln [1+e^{-\beta \left(
\epsilon _{\vec{k},\alpha ,n}-\lambda \right) }]+\lambda +\frac{t^{2}}{2}%
\left( 1+r^{2}\right) ,
\]
where $\beta =1/k_{B}T$, and $\lambda $ is determined by the electron number
eq.,
\[
\frac{1}{N}\sum_{\vec{k},\alpha ,n}\frac{1}{1+e^{\beta \left( \epsilon _{%
\vec{k},\alpha ,n}-\lambda \right) }}=1.
\]
The mean field amplitudes $t$ and $r$ are obtained by minimizing the free
energy at a given temperature. As $T$ is increased from zero, we find that
the $F$ vs. $r$ curve is qualitatively identical to the $E$ vs. $r$ curve at $T=0
$ (Fig.(\ref{fig2})). Thus, the disconnected plaquette state ($r=1$) remains
as the lowest free energy state (note that since $r=1$ is trivially a
self-consistent solution, it is always a free energy extremum). Meanwhile
the short-ranged correlation $t$ decreases as $T$ increases. In Fig.(\ref
{fig3}) we show the values of $t$ that minimizes the free energy for $r=0$
and $r=1$ as a function of $T.$ Other $r$ value curves lie between these two
curves. We see that at critical temperature $T_{t}=0.75,$ $t\rightarrow 0$
for all values of $r$. Above $T_{t},$ even nearest neighbor sites are
uncorrelated. $T_{t}$ can be calculated analytically easily. The free energy
$F_{MF}$ for $H_{MF}$ in eq.(\ref{eqn7}) can be formally expanded in a series in $%
t_{ij}.$ Assuming the transition in $t$ is continuous (which is likely since $%
F_{MF}$ is even in $t)$, $T_{t}$ is given by the temperature where the
coefficient in front of the quadratic term changes sign from positive to
negative. To quadratic order, $F_{MF}$ can be easily found using
perturbation by taking the chemical potential part of $H_{MF}$ as the
unperturbed Hamiltonian $H_{0}$ and the remaining part as the perturbation $%
H^{\prime }$(essentially doing high temperature expansion in $H^{\prime }$).
By expanding to second order in $H^{\prime },$ we  obtain
\begin{eqnarray}
F_{MF}=F_{0}+\sum_{\left\langle ij\right\rangle }\left( 1-\frac{3}{4}\beta
\right) t_{ij}^{2}+{\it O}(t^{4}).
\label{eqn7}
\end{eqnarray}
 From this, we see that $T_{t}=3/4$ and that $T_{t}$ is independent of the
configuration of $t_{ij},$ hence of $r.$ The best configuration for $t_{ij}$
is determined by the higher order terms in $F_{MF}.$ Our calculation
indicates it corresponds to $r=1.$ Since for all $T<T_{t}$, $r=1$, the
transition in $r$ is preempted by the transition in $t.$ Within our MFT, we
are therefore not able to answer the question of whether the plaquette
ordering transition is a first order or a continuous one. However, the mean
field transition in $t$ is unphysical and merely signifies a crossover from
weak to strong nearest neighbor correlations. We thus interpret our mean
field result to imply that as soon as significant nearest neighbor
correlations develop, the system has complete plaquette ordering.

We now turn to the triangular lattice case. Similar to the square lattice,
we consider plaquette ordering due to spontaneous formation of different
(type-A and type-B) bond amplitudes as shown in Fig.(\ref{fig1}). On the
triangular lattice, there are more type-B bonds than A-type. Out of 3N
bonds, 5N/4 are A-type, and 7N/4 are B-type. Rigorously speaking, the bonds
within A-type (or B-type) are not all equivalent. For example, among the
A-type bonds illustrated in Fig.(\ref{fig1}), the diagonal one has different
connectivity from the four edge bonds. This requires, in principle, to
assign different amplitudes for these bonds in a fully self-consistent mean
field theory. Instead, we approximate all A bonds and all B bonds to be
identical among themselves, which simplifies the calculation greatly. Our
calculation then amounts to a variational calculation with two types of
bonds or equivalently, a mean field calculation that is self-consistent when
averaged over the A bonds or over the B bonds. The physics should not be
affected qualitatively by this simplification. Quantitatively, we have also
checked this point by examining a single A-plaquette problem explicitly.
Assuming all bonds to be equivalent, we find $t_{a}=1.025,$ while allowing
the $4$ edge bonds ($t_{a}^{\prime })$ and the diagonal bond ( $%
t_{a}^{\prime \prime })$ to be different, we find $t_{a}^{\prime }=0.943$
and $t_{a}^{\prime \prime }=1.333,$ giving an average of $1.021.$ The ground
state energies in the two cases are also within $2\%$ of each other, with
values $-1.31$ and $-1.33$ respectively. The simplified scheme of only
allowing two different bond strengths is therefore sufficient for our
purpose.

We thus introduce two average mean field amplitudes: one for A-type and one
for B-type. The self-consistent condition is then given by their average
one, $t_{a}=\sum \langle \Phi _{ij}\rangle /N_{A}$, and $t_{b}=\sum \langle
\Phi _{ij}\rangle /N_{B}$, where $\langle {i,j}\rangle \in \{A\}$ and $\in
\{B\}$ respectively, and $N_{A}$ and $N_{B}$ are the number of A and B type
bonds. With this approximation, the mean field Hamiltonian in a triangular
lattice reads,

\begin{eqnarray}
H_{mf}= &-&\sum_{\vec{k},n}[t\,\left( \cos {k_{x}a}+\cos {k_{y}a}\right)
+\left( 1-r/2\right) t\,\cos {\left( k_{x}a+k_{y}a\right) }]~c_{\vec{k}%
,n}^{\dagger }c_{\vec{k},n}  \nonumber \\
&+&ir\,t\,\sum_{\vec{k},n}\left( \sin {k_{x}a}+\frac{1}{2}\sin {\left(
k_{x}a+k_{y}a\right) }\right) c_{\vec{k}+\left( \pi ,0\right) ,n}^{\dagger
}c_{\vec{k},n}  \nonumber \\
&+&ir\,t\,\sum_{\vec{k},n}\left( \sin {k_{y}a}+\frac{1}{2}\sin {\left(
k_{x}a+k_{y}a\right) }\right) c_{\vec{k}+\left( 0,\pi \right) ,n}^{\dagger
}c_{\vec{k},n}  \nonumber \\
&-&r\,t\,\sum_{\vec{k},n}\frac{1}{2}\cos {\left( k_{x}a+k_{y}a\right) }~c_{%
\vec{k}+\left( \pi ,\pi \right) ,n}^{\dagger }c_{\vec{k},n}  \nonumber \\
&+&\frac{Nt^{2}}{4}\left( 3+3r^{2}-r\right) \,-\lambda \left( \sum_{\vec{k}%
,n}~c_{\vec{k},n}^{\dagger }c_{\vec{k},n}-N\right) .
\label{eqn8}
\end{eqnarray}
In the above eqn., we have used the same notation as for the square lattice.
Note that in the triangular lattice, there is a lack of symmetry between the
strong and the weak bonds. Consequently, the Hamiltonian does not possess
all the symmetries of the square lattice. In particular, the system is not
symmetric under the transformation $r\rightarrow -r$. From the definition of
$r$ and $t$, we have a symmetry in parameter space, $r\rightarrow
1/r,t\rightarrow r\,t$. This allows us to study $r$ within [-1, 1]. $%
t\rightarrow -t$ symmetry is also not present for the triangular lattice.
For $r=0$ and $r=1,$ the energy is minimized by $t$ positive, and we assume
this holds for other $r\neq 0$ also. The mean field Hamiltonian can be written in
the diagonalized form,

\[
H_{mf}=\sum_{\vec{k},\alpha ,n}\epsilon _{\vec{k},\alpha ,n}\beta _{\vec{k}%
,\alpha ,n}^{\dagger }\beta _{\vec{k},\alpha ,n}+\frac{Nt^{2}}{4}[3\left(
1+r^{2}\right) -r]-\lambda \left( \sum_{\vec{k},\alpha ,n}\beta _{\vec{k}%
,\alpha ,n}^{\dagger }\beta _{\vec{k},\alpha ,n}-N\right).
\]
The ground state is given by a filled Fermi sea , whose total energy is
given by

\begin{eqnarray}
\frac{E}{N}=\frac{1}{N}\sum_{\vec{k},\alpha ,n}~\epsilon _{\vec{k},\alpha
,n}\theta \left( \lambda -\epsilon _{\vec{k},\alpha ,n}\right) +\frac{t^{2}}{%
4}[3\left( 1+r^{2}\right) -r].  \label{eqn9}
\end{eqnarray}
Since there is no $r \rightarrow -r$
symmetry,  $E(r)\neq E(-r).$ Our mean field result shows that $E(r)$
has lower energy for positive $r.$ Similar to the square lattice, the ground
state is the disconnected plaquette state corresponding to $r=1$. In this
case, the ground state is $12$-fold degenerate. Again, the local stability
of $r=1$ can be confirmed by second order perturbation theory. This also
gives a heuristic understanding of why the energy is lower for positive vs.
negative $r.$ The $r=-1$ state is a plaquette state such that neighboring
plaquettes are connected by a single diagonal bond of equal strength as the
bonds in the plaquette. Now imagine this connecting bond is not equal but
much weaker, which will make this state quite similar to $r$ less than but
very close to $1.$ The local stability of $r=1$ is thus consistent with it
being lower in energy than $r=-1,$ and by continuity with positive $r$
having the lower energy vs. negative $r$ for at least a range of $r$ close to
$1.$

\begin{figure}[p]
\epsfxsize=4.5in
\begin{center}
\epsffile{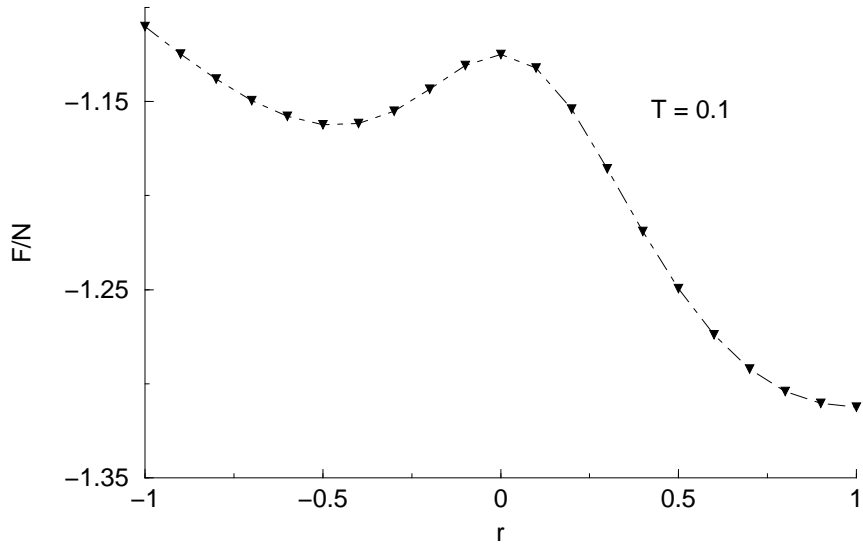}
\end{center}
\caption{{{Free energy per site from eqn.(\ref{eqn10}) is plotted
with respect to $r$ for the triangular lattice case at $T = 0.1$.
At this temperature, $r=1$ remains the stable state and $r=0$ remains as a local maximum.
}}}
\label{fig4}
\end{figure}

\begin{figure}[p]
\epsfxsize=6.5in
\begin{center}
\epsffile{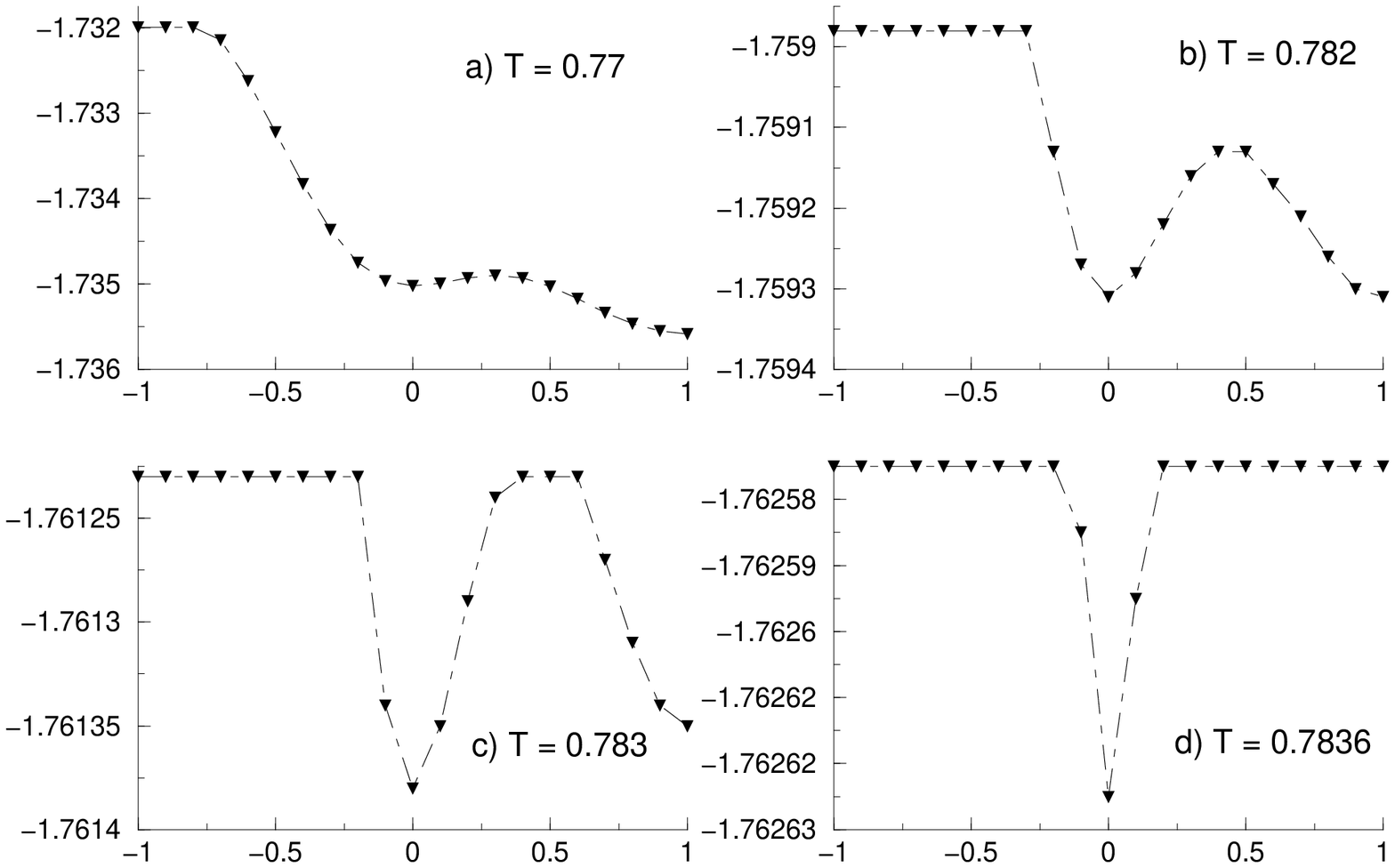}
\end{center}
\caption{{{Free energy per site v.s.\,  $r$ on the triangular lattice.
a) For $0.74 < T < 0.782$, $r=1$ is the lowest free energy state while $r=0$ is
locally stable. \, b) At $T=0.782$, $r=0$ and $r=1$ become degenerate. At this T, the transition
in $t$ for a given $r$  has already occurred for a range of $r < 0$. \, c) At T=0.783,
$r=0$ is the stable state and $r=1$ is a local minimum. Transition in $t$
begins to occur for some $r > 0$.
d) At $T=0.7836$, the transition in $t$ has occurred for $r=1$, leaving $r=0$ as
the only minimum.
}}}
\label{fig5}
\end{figure}

To study the finite temperature phase, we minimize the free energy

\begin{eqnarray}
\frac{F}{N} = - \frac{k_{b}T}{N} \sum_{\vec{k},\alpha,n} \ln [ 1 +
e^{-\beta\left(\epsilon_{\vec{k},\alpha,n} - \lambda \right)}] + \lambda &+&
\frac{ t^{2}}{4}[3\left(1+r^{2}\right) - r],  \nonumber \\
\frac{1}{N} \sum_{\vec{k},\alpha,n} \frac{1}{1+e^{ \beta\left(\epsilon_{\vec{%
k},\alpha,n} - \lambda \right)}} &=& 1.  \label{eqn10}
\end{eqnarray}
As in the square lattice, $r=1$ remains the stable state at low $T$ 
(Fig.(\ref{fig4})).
On the
triangular lattice however, the transition in $r$ is not preempted by the
one in $t,$ and we are able to study its phase transition$.$ At $T\approx %
0.74,$ the curvature of the free energy curve at $r=0$ changes sign and $r=0$
changes from a local maximum to local minimum. When $T$ is increased to $%
T_{c}=0.782,$ $r=0$ and $r=1$ states become degenerate in free energy and
there is a first order transition from the disconnected plaquette state to
the uniform state (or the other way round if we consider lowering the
temperature). These results are summarized in Fig(\ref{fig5}).
Since the lack of reflection symmetry in $r$ allows for a
cubic invariant in the Landau theory for its phase transition, the first
order transition obtained is to be expected. Indeed, the behavior of $F$ vs.
$r$ as a function of $T$ is precisely what one would expect from such a
Landau-Ginzburg free-energy functional. Upon further raising the temperature,
the average nearest neighbor correlation continues to decrease, with the
mean field transition to a completely uncorrelated state occurring at $T_{t}%
\approx 0.8.$ Unlike the square lattice case, on the triangular lattice, the
transition in $t$ occurs at different temperature for different $r,$ i.e. $%
T_{t}=T_{t}(r).$  This can be readily understood from the lack of $%
t\rightarrow -t$ symmetry. As a result, the corresponding $F_{MF}$
(eq.(\ref{eqn7}))
contains odd power terms, which in turn results in first order transition in
$t_{ij}.$ Naturally then, the transition temperature will occur above the
second order temperature at $0.75$ and will depend on the configuration of $%
t_{ij,}$ hence on $r.$ As always, the transition in $t$ should be
interpreted physically as a crossover.

In summary, within MFT, we find the uniform mean field state to be unstable
with respect to plaquette ordering at $T=0$ for both the square and
triangular lattices. For the triangular lattice, the plaquette ordering
transition at finite temperature is first order, in agreement with Landau
theory argument. For the square lattice, the plaquette to uniform state is
preempted by a mean field transition in $t,$ and MFT does not give an
unambiguous answer to if the plaquette transition is first order or
continuous. Taking the MFT on its face value, the transition is continuous as
there is no entropy jump, although this is complicated by it being also a
transition in $t.$ Based on the reflection symmetry in $r,$ Landau theory
would also suggest a continuous transition. However, we cannot rule out a
first order transition due to a negative quartic invariant in the Landau
free energy functional or due to fluctuations. For the triangular lattice,
the plaquette transition is first order as expected from symmetry.

Within MFT, the ground states of the SU(4) model in both square and
triangular lattice are found to be disconnected plaquette states ($r=1$).
Naively, this implies huge degeneracies in the ground states. Consider a
lattice with periodic boundary condition. All the plaquette states with
plaquettes covering all the lattice sites have the same lowest mean field
energy. A simple lower bound for ground state degeneracy would then be $%
D\approx 2^{\sqrt{N}}$ for the square lattice, and $D\approx 3^{N/12}$ for
the triangular lattice. However, such plaquette states are not orthogonal to
each other whether we consider the mean field states or the actual SU(4)
states. They may not even be independent, so these lower bounds for
degeneracy may not be valid, but should be taken as suggestive of huge
degeneracy. Obviously, the exact SU(4) ground state will not exhibit
saturated plaquette ordering (there will be for example resonance between
plaquette states), and the actual ground state will only be $4$-fold and $12$%
-fold degenerate respectively. Nevertheless, we may still interpret the mean
field ground state to imply a very strong, although not saturated plaquette
ordering, so that plaquettes are only weakly coupled. We would then expect a
huge number of low lying energy states in the energy spectrum. This is
consistent with the preliminary results of Penc et al.~\cite{penc} on the
square and triangular lattices using variational calculations on relatively
large systems.

This work was supported in part by DOE grant No. DE/FG03-01ER45687 and by
URC summer fellowship at University of Cincinnati. The authors wish to thank
F. Mila and X. R. Wang for discussions.



\end{document}